\documentclass[preprint,amsmath,amssymb,superscriptaddress,notitlepage]{revtex4-1}
\usepackage{epsfig}
\usepackage{graphicx}
\usepackage{color}
\usepackage{bm}

\begin{document}

\title{Large anomalous Nernst effect in thin films of the Weyl semimetal Co$_2$MnGa}

\author{Helena~Reichlova}
\affiliation{Institut f{\"u}r Festk{\"o}rper- und Materialphysik, Technische Universit{\"a}t Dresden, 01062 Dresden, Germany}
\affiliation{Institute of Physics ASCR, v.v.i., Cukrovarnick\'a 10, 162 53, Praha 6, Czech Republic}
\author{Richard Schlitz}
\affiliation{Institut f{\"u}r Festk{\"o}rper- und Materialphysik, Technische Universit{\"a}t Dresden, 01062 Dresden, Germany}
\affiliation{Center for Transport and Devices of Emergent Materials, Technische Universit{\"a}t Dresden, 01062 Dresden, Germany}
\author{Sebastian~Beckert}
\affiliation{Institut f{\"u}r Festk{\"o}rper- und Materialphysik, Technische Universit{\"a}t Dresden, 01062 Dresden, Germany}
\author{Peter~Swekis}
\affiliation{Institut f{\"u}r Festk{\"o}rper- und Materialphysik, Technische Universit{\"a}t Dresden, 01062 Dresden, Germany}
\affiliation{Max Planck Institute CPfS, Dresden, Germany 01187}
\author{Anastasios~Markou}
\affiliation{Max Planck Institute CPfS, Dresden, Germany 01187}
\author{Yi-Cheng Chen}
\affiliation{Max Planck Institute CPfS, Dresden, Germany 01187}
\affiliation{Department of Materials Science and Engineering, National Chiao Tung University, Hsinchu City, Taiwan 30010}
\author{Savio~Fabretti}
\affiliation{Institut f{\"u}r Festk{\"o}rper- und Materialphysik, Technische Universit{\"a}t Dresden, 01062 Dresden, Germany}
\author{Gyu Hyeon Park}
\affiliation{Leibniz Institute for Solid State and Materials Research Dresden (IFW Dresden), Institute for Metallic Materials, 01069 Dresden, Germany}
\author{Anna Niemann}
\affiliation{Leibniz Institute for Solid State and Materials Research Dresden (IFW Dresden), Institute for Metallic Materials, 01069 Dresden, Germany}
\author{Shashank Sudheendra}
\affiliation{Leibniz Institute for Solid State and Materials Research Dresden (IFW Dresden), Institute for Metallic Materials, 01069 Dresden, Germany}
\author{Andy Thomas}
\affiliation{Leibniz Institute for Solid State and Materials Research Dresden (IFW Dresden), Institute for Metallic Materials, 01069 Dresden, Germany}
\affiliation{Center for Transport and Devices of Emergent Materials, Technische Universit{\"a}t Dresden, 01062 Dresden, Germany}
\author{Kornelius Nielsch}
\affiliation{Leibniz Institute for Solid State and Materials Research Dresden (IFW Dresden), Institute for Metallic Materials, 01069 Dresden, Germany}
\affiliation{Technische Universit{\"a}t Dresden, Institute of Materials Science, 01062 Dresden, Germany}
\author{Claudia~Felser}
\affiliation{Max Planck Institute CPfS, Dresden, Germany 01187}
\author{Sebastian T. B. Goennenwein}
\affiliation{Institut f{\"u}r Festk{\"o}rper- und Materialphysik, Technische Universit{\"a}t Dresden, 01062 Dresden, Germany}
\affiliation{Center for Transport and Devices of Emergent Materials, Technische Universit{\"a}t Dresden, 01062 Dresden, Germany}

\begin{abstract}
The magneto-thermoelectric properties of Heusler compound thin films are very diverse. Here, we discuss the anomalous Nernst response of Co$_2$MnGa thin films. We systematically study the anomalous Nernst coefficient as a function of temperature, and we show that unlike the anomalous Hall effect, the anomalous Nernst effect in Co$_2$MnGa strongly varies with temperature. We exploit the on-chip thermometry technique to quantify the thermal gradient, which enables us to directly evaluate the anomalous Nernst coefficient. We compare these results to a reference CoFeB thin film. We show that the 50-nm-thick Co$_2$MnGa films exhibit a large anomalous Nernst effect of -2~$\mu$V/K at 300~K, whereas the 10-nm-thick Co$_2$MnGa film exhibits a significantly smaller anomalous Nernst coefficient despite having similar volume magnetizations. These findings suggest that the microscopic origin of the anomalous Nernst effect in Co$_2$MnGa is complex and may contain contributions from skew-scattering, side-jump or intrinsic Berry phase. In any case, the large anomalous Nernst coefficent of Co$_2$MnGa thin films at room temperature makes this material system a very promising candidate for efficient spin-caloritronic devices.   

\end{abstract}


\maketitle
\pagebreak

 Heusler compounds exhibit various different properties and phenomena, which makes them a very interesting class of materials \cite{Graf2011,Wollmann2017}. For example, by selecting the appropriate Heusler compound, one can make the material metallic, semimetallic, half-metallic, or semiconducting. Likewise, Heusler compounds are known to exhibit many different types of magnetic order \cite{Sanvito2017}, and the strength of the spin-orbit interaction and related phenomena can be engineered. This wide range of properties within one family of materials enables one to design multilayered systems with structurally and chemically compatible interfaces and notably different physical properties, which consequently enable efficient spin injection \cite{Chadov2011} or exchange coupling \cite{Ranjbar2015}.
In particular, some Co$_2$YZ-based full Heusler compounds have high Curie temperature and high spin polarization, which result in a tunneling magnetoresistance (TMR) above 500$\%$ \cite{Ikeda2007}.
Last but not least, Heusler compounds offer a notably promising route towards magnetic Weyl semimetals \cite{Felser2016}. Weyl semimetals can potentially host unique transport phenomena because of the nontrivial topology of the band structure \cite{Kuebler2016,Chang2016,Shi2018}, which also makes the spin-caloritronic response of Weyl semimentals very interesting \cite{Bosu2011,Yamasaki2015,Boehnke2017,Noky2018}. One of the key effects in spin-caloritronic research is the anomalous Nernst effect (ANE), which is the thermal counterpart of the anomalous Hall effect (AHE). The ANE is experimentally observed as a transverse voltage generated in a magnetic material subjected to a thermal gradient. The ANE was considered to be proportional to magnetization \cite{Smith1911,Pu2008}, but recent studies suggest that Berry curvature effects \cite{Xiao2006,Ikhlas2017} also can play a dominating role. 

Here, we focus on Co$_2$MnGa, which is an ideal representative of the aforementioned Heusler materials. Co$_2$MnGa is considered half metalic with a high Curie temperature T$_C\sim$700~K \cite{Webster1970}. Recently, Co$_2$MnGa was suggested as an ideal candidate for the experimental study of unconventional topological surface states \cite{Chang2017,Kuebler2016}. Despite the promising properties of Co$_2$MnGa, not much is known about the details of its electronic properties \cite{Kurtulus2005,Manna2017}. Even fewer studies are available on the thin-film fabrication and characterization of this material \cite{Ludbrook2017,Pechan2005}. In the works of Ludbrook et al. and Markou et al., the anomalous Hall effect was systematically measured in Co$_2$MnGa thin films, which yield nontrivial temperature and thickness dependency \cite{Ludbrook2017,Markou2018u}. The authors suggested that various contributions, i.e., skew-scattering, side-jump scattering and the intrinsic Berry phase effects, contribute to the AHE in Co$_2$~-~based Heusler compounds.
The magneto-thermoelectric properties of Co$_2$MnGa have not been studied, yet. In this manuscript, we study the anomalous Nernst response of Co$_2$MnGa thin films. We show that Co$_2$MnGa thin films exhibit a high anomalous Nernst coefficient $N_{ANE}\sim$~-2$\mu$V/K at room temperature despite their moderate saturation magnetization M$_{sat}\sim$700~kA/m. Since the anomalous Nernst coefficient in a conventional 3d transition metal thin film is typically smaller than 1$\mu$V/K \cite{Mizuguchi2012,Sakuraba2013,Hasegawa2015,Chuang2017,Ramos2013,Uchida2015,Ando2018}, our results suggest that the nontrivial topology of band structure plays an important role for the magneto-thermogalvanic response of Co$_2$MnGa. By comparing the anomalous Nernst response of Co$_2$MnGa and reference CoFeB thin films, we demonstrate that the ANE is not directly proportional to the absolute value of the saturation magnetization. Furthermore, we observe that $N_{ANE}$ is reduced in thin Co$_2$MnGa films and discuss possible reasons for this effect.

Co$_2$MnGa films with two different thicknesses of 10 and 50~nm were deposited by magnetron sputtering on single-crystal MgO (100) substrates using a multisource Bestec UHV deposition system.
 The Co (5.08~cm diameter) and Mn$_{50}$Ga$_{50}$ (5.08~cm diameter) sources were in confocal geometry, and the target-to-substrate distance was 20~cm. Prior to the deposition, the chamber was evacuated to a base pressure below 2x10$^{-8}$ mbar, and the process gas (Ar) pressure during deposition was 3x10$^{-3}$ mbar. The Co$_2$MnGa films were grown by cosputtering. Co and MnGa were deposited by applying 34~W and 22~W dc power, respectively. The growth rates and film thicknesses were determined using a quartz crystal microbalance and confirmed by X-ray reflectivity measurements. The substrate was rotated during deposition to ensure homogeneous growth. The Co$_2$MnGa films were grown at 500$^{\circ}$C and postannealed in situ for an additional 20 min at 500$^{\circ}$C. All samples were capped at room temperature with a 2-nm-thick Al film to prevent oxidation. The stoichiometry was estimated by Energy-dispersive X-ray spectroscopy and verified by inductively coupled plasma optical emission spectrometry. The high structural quality was confirmed with X-ray diffraction.

Reference samples of the well-studied metallic ferromagnet CoFeB (15 and 60~nm) were prepared on Si/SiO substrates.
The magnetization of all samples was measured in a Quantum Design SQUID magnetometer. The results are shown in Figs.~1a and b: the similar volume magnetization of $\sim$700~kA/m at room temperature for the two materials is consistent with that of previous studies \cite{Yamazoe2016,Ludbrook2017}. 
The thin films were then patterned into Hall bar by optical lithography and plasma etching. A microscope image of the Hall bar is shown in Fig.~2a. The layout of the Hall bar has long transverse contacts to maximize the ANE signal, since the electric field generated by the ANE is proportional to the distance between transversal contacts.

\begin{figure}[h]
\hspace*{0cm}\epsfig{width=1\columnwidth,angle=0,file=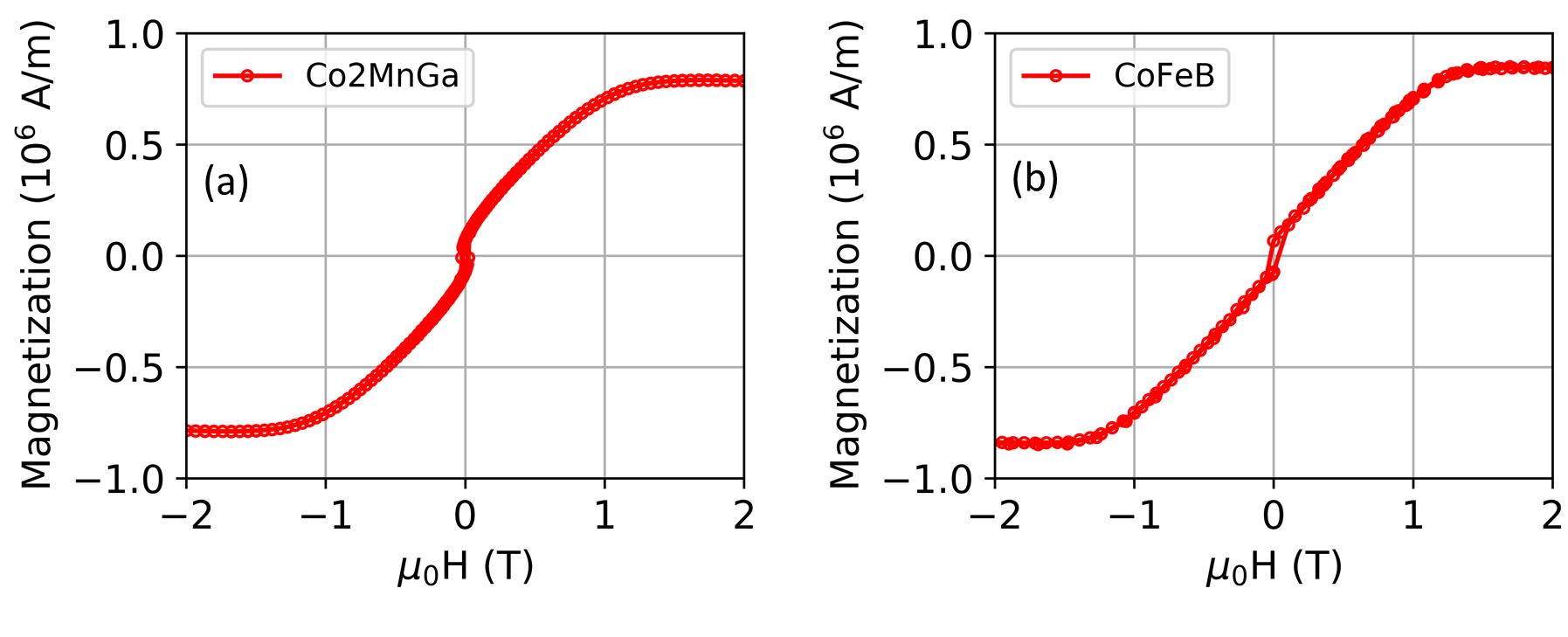}
\caption{ Sample characterization. (a),(b) Magnetization of the Co$_2$MnGa 50~nm (a) and reference CoFeB 15~nm (b) films measured using SQUID magnetometry at 300~K with the magnetic field perpendicular to the sample plane.}
\label{f11}
\end{figure}

A schematic picture of the sample and experiment geometry is shown in Fig.~2b. The sample was glued on two macroscopic blocks made from brass and plastic to ensure notably different thermal couplings between the substrate and the block at either side of the sample. A thermal gradient $\nabla$T was generated in the sample plane by a Pt heater placed on the downside of the substrate (the "hot" side is shown in red in Fig.~2b). An external magnetic field was applied perpendicular to the sample plane. The voltage $\text{V}_{ANE}$ was detected on a transversal pair of contacts, and the thermal gradient was quantified using on-chip thermometry as detailed below. Multiple pairs of contacts were used to detect $\text{V}_{ANE}$ under the same thermal gradient, which resulted in a voltage variation below 10~$\%$, confirming that the thermal gradient is linear in the sample plane. 

\begin{figure}[h]
\hspace*{0cm}\epsfig{width=1\columnwidth,angle=0,file=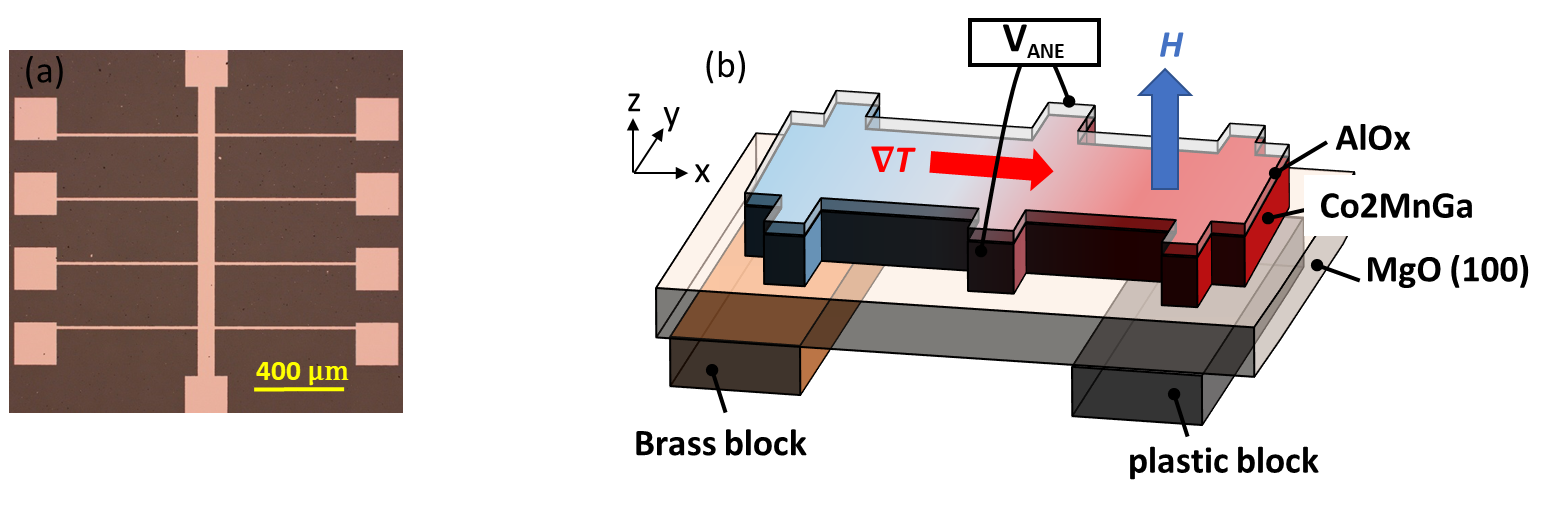}
\caption{ Experimental setup. (a) Microscope image of the device design. The Hall bar was defined by optical lithography; the bright areas correspond to Co$_2$MnGa, whereas the dark areas indicate the bare substrate (Co$_2$MnGa removed by etching). (b) Schematics of the experimental setup (not to scale). A thermal gradient was applied in the plane of the sample and the magnetic field was applied out of the sample plane. The Nernst voltage $\text{V}_{ANE}$ was measured on a pair of transversal contacts. }
\label{f12}
\end{figure}

All presented transport data were measured in an Oxford Instruments cryostat with a variable-temperature insert equipped with two thermometers to monitor the sample base temperature with high precision. We use standard Keithley 2182 nanovoltmeters to detect $\text{V}_{ANE}$. In addition, to independently record the AHE response of our samples, a DC current was applied along the Hall bar, and the transversal voltage was measured as a function of the out-of-plane magnetic field. We obtain an AHE with anomalous Hall conductivity $\sigma_{AHE}$ = 600 $\Omega.cm^{-1}$, which is consistent with Markou et al.\cite{Markou2018u}. Furthermore, the AHE is almost independent of temperature in the studied temperature range of 10~K$ < $~T $< $~300~K and shows only a slight increase from  $\sim$600~$\Omega.cm^{-1}$ to 500~$\Omega.cm^{-1}$ with decreasing temperature.

To quantitatively determine the ANE coefficient, quantitative knowledge of the thermal gradient $\nabla T$ (i.e., the temperature difference $\Delta T$ per length along the sample) is mandatory. Therefore, we focused on the evaluation of $\nabla T$ in our structures. A first rough estimate of $\Delta T$ between the hot and cold parts of the sample is derived using two Pt thermometers, which were placed on the plastic and brass blocks below the substrate. However, this estimate of $\Delta T$ in our thin film is inaccurate because the thermal contact resistance of the heater, glue and substrate must be considered (see the schematic sample mounting in Fig.~2b). Thus, we developed the following measurement algorithm to directly obtain $\Delta T$ "on chip" in the measured structure.
The on-chip thermometry setup is schematically shown in Fig.~3a. As small probing current of 100~$\mu$A was applied along the transverse contact arms, and the voltages V$_1$ and V$_2$ on two ends of the Hall bar were measured. First, the heating power is set to zero ($\Delta T = 0$), and the base temperature of the entire setup is homogeneously heated using the VTI and sample rod heaters to accurately control the temperature. Hence, we obtain the calibration curves V$_1$(T)/I and V$_2$(T)/I that connect the resistance of the transverse arms with the (local) film temperature. By fitting the two curves with a polynomial fit, we obtain the temperature calibration for the two pairs of contacts. The measured voltages V$_1$ and V$_2$ as functions of temperature are shown in Fig.~3b. The heater power is then gradually increased, and voltages $V_1$ and $V_2$ varied, as shown in Figs.~3c and d for one particular base temperature T~=~300~K. Using the fitted curves, we determined the temperature at the hot and cold ends of the sample and consequently evaluated $\nabla T$. This procedure is repeated at each base temperature where ANE measurements were performed to yield a typical thermal gradient of $\nabla T \sim$~0.4~K/mm. At temperatures above 100~K, the on-chip thermometry yields a robust number for $\nabla T$ with a resolution of 0.1~K/mm. At lower temperatures (below 100~K), the nonlinearity of the calibration curves increases the error in the fit and reduces the resolution to 0.3~K/mm. Figure~3a shows a schematic; in the real device (Fig.~2a), the main voltage drop is along the long transversal contacts, which results in a well-defined position of the temperature with respect to the thermal gradient. 

\begin{figure}[h]
\hspace*{0cm}\epsfig{width=1\columnwidth,angle=0,file=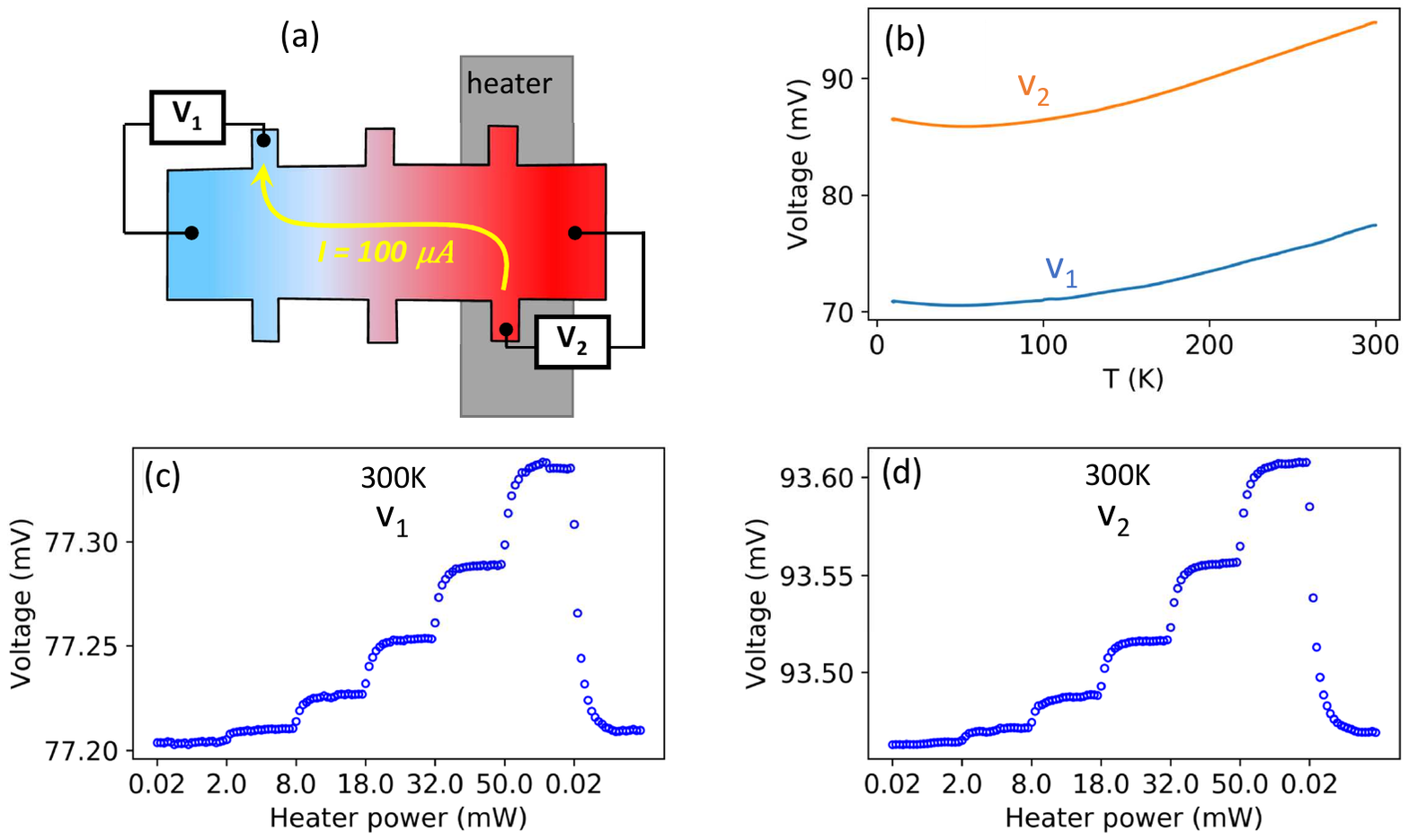}
\caption{ Thermal gradient evaluation. (a) Schematic picture of the setup. (b) Calibration V$_1$(T) and V$_2$(T) curves measured at the "cold" and the "hot" pairs of contacts, respectively, while homogeneously heating the sample (no thermal gradient applied). (c), (d) Voltages V$_1$ and V$_2$ as functions of the heater power measured for a global temperature of 300~K. }
\label{f2}
\end{figure}

The results of the ANE measurements are presented in Fig.~4. The ANE experiments were performed by first adjusting the global cryostat temperature, then applying $\nabla T$, measuring the local temperature and finally measuring the thermovoltage $\text{V}_{ANE}$. Figures~4a, b, and c show $\text{V}_{ANE}$ as a function of the magnetic field measured at various temperatures for the 50-nm-thick Co$_2$MnGa sample, 10-nm-thick Co$_2$MnGa samples and CoFeB reference samples, respectively. As a reference, we show the data measured on the 15-nm-thick CoFeB film. However, we have confirmed that 60-nm-thick CoFeB in a similar configuration yields identical values of the anomalous Nernst coefficient. The measured voltage saturates at 1.2~T, which corresponds to the out-of-plane magnetic anisotropy of our material as shown in Figs.~1a and b. After the saturation, only a weak variation of the measured voltage is observed. Therefore, we conclude that our signal is dominated by the ANE with negligible contribution from the ordinary Nernst effect. 
The anomalous Nernst coefficient $N_{ANE}$ is evaluated as follows:
\begin{equation}
{E}_{ANE} = - N_{ANE} {\bf m}\times{\bf \nabla T},
\end{equation}
where ${\bf m}$ is the normal vector along the magnetization.

The obtained anomalous Nernst oefficients for all three samples are shown in Fig.~4d. The measured $N_{ANE}$ is $\sim$~ -2 $\mu$V/K at 300~K in the 50-nm-thick Co$_2$MnGa film, which is large compared to those of other thin ferromagnetic films, whose typical values of $\lvert N_{ANE}\rvert$ are below 1~$\mu$V/K \cite{Hasegawa2015,Chuang2017,Ramos2013,Uchida2015,Ando2018,Ikhlas2017}. The magnitude of the measured $N_{ANE}$ is even more significant when we consider that in ferromagnetic metals, $N_{ANE}$ generally increases with the saturation magnetization as shown in \cite{Ikhlas2017,Hasegawa2015}. Co$_2$MnGa has a moderate saturation magnetization of $\sim$700~kA/m, but the measured magnitude of $N_{ANE}$ is comparable to that of materials that are known for having higher saturation magnetizations \cite{Slachter2011}. This observation is even more striking when we compare our results measured on Co$_2$MnGa and the reference CoFeB films measured under identical conditions. The samples have similar magnetizations, but $N_{ANE}$ of CoFeB is $\sim$~ -0.8~$\mu$V/K, which is consistent with the general trend in \cite{Ikhlas2017}, unlike Co$_2$MnGa. 
 
The magnitude of $N_{ANE}$ decreases with decreasing temperature, which is in contrast to the temperature depedence of the AHE. This dependence is consistent with previous studies \cite{Miyasato2007,Gautam2018} and is generally understood as a consequence of Mott relation \cite{Xiao2006}. At low temperatures, $N_{ANE}$ is expected to linearly increase with increasing temperature, whereas $N_{ANE}$ is dominated by the changes in the band structure at the Fermi level near Curie temperature $T_C$. Since $T_C$ of Co$_2$MnGa is far above the studied temperature range \cite{Webster1970}, the presented data are expected to linearly increase with increasing $T$. This behavior is observed for the reference CoFeB sample and 10-nm-thick Co$_2$MnGa sample. However, $N_{ANE}$ measured in the 50-nm-thick Co$_2$MnGa film does not follow the linear trend, which calls for further studies. 

Figure~4d shows that the magnitude of $N_{ANE}$ is significantly reduced in the thin Co$_2$MnGa film. The variation in $N_{ANE}$ with the thickness of the Co$_2$MnGa film may be caused by several factors in principle. First, the thicker Co$_2$MnGa film can have different structural \cite{Claydon2007} or magnetic properties \cite{Ludbrook2017}. However, the magnetometry does not show a significant difference in volume magnetization for the two samples. The X-ray characterization did not reveal any change of structure between the two samples. A more likely scenario is that similar to the AHE \cite{Ludbrook2017}, various intrinsic and extrinsic contributions affect the magnitude of $N_{ANE}$. This scenario is supported by the reported nontrivial topology of Co$_2$MnGa \cite{Chang2017,Kuebler2016}, the large $N_{ANE}$ measured in the 50-nm-thick Co$_2$MnGa film and its nonlinear increase with increasing temperature. Further study of the thickness dependency of Co$_2$MnGa should be performed but is out of the scope of this manuscript. 

In summary, we have measured for the first time the anomalous Nernst effect in the promising Heusler compound Co$_2$MnGa. We evaluated the anomalous Nernst coefficient $N_{ANE}$ at various temperatures and observe that unlike the anomalous Hall conductivity, $N_{ANE}$ increases with increasing temperature and does not saturate until 300~K, which makes the material appealing for new spin-caloritronic devices \cite{Sakuraba2013,Narita2017}. The magnitude of $N_{ANE}$ $\sim$~-2 $\mu$V/K at 300~K in the 50-nm-thick Co$_2$MnGa film is large compared to that reported in other ferromagnetic thin films. A reference CoFeB sample exhibits an $N_{ANE}$ value that is smaller by a factor of 2, although CoFeB and Co$_2$MnGa have similar saturation magnetization. Moreover, the magnitude of $N_{ANE}$ in Co$_2$MnGa strongly varies with the film thickness, which suggests that a subtle interplay of intrinsic and extrinsic contributions must be considered to describe the ANE in this material. Our study contributes to the understanding of the spin-dependent thermoelectric transport properties in the very promising material Co$_2$MnGa.

\begin{figure}[h]
\hspace*{0cm}\epsfig{width=1\columnwidth,angle=0,file=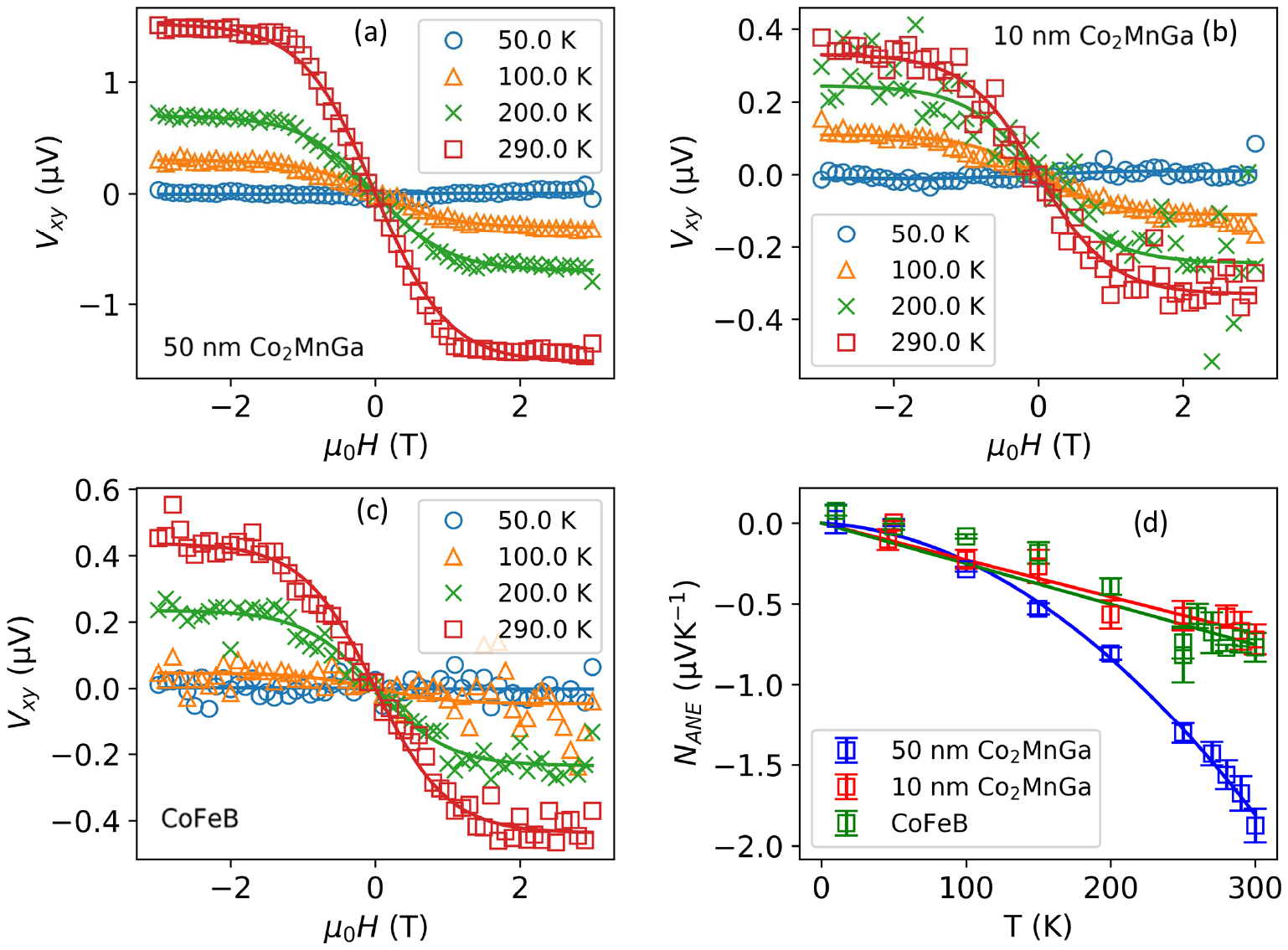}
\caption{ Anomalous Nernst measurement. $V_{ANE}$ as a function of the measured out-of-plane magnetic field in Co$_2$MnGa 50~nm (a), Co$_2$MnGa 10~nm (b) and reference CoFeB (c) at four different temperatures. Panel (d) shows the anomalous Nernst coefficient evaluated of the 3 samples as a function of temperature. }
\label{f3}
\end{figure}

\pagebreak

We acknowledge Dominik Kriegner for useful comments on data visualisation. We acknowledge funding via the priority programme Spin Caloric Transport (spinCAT) of Deutsche
Forschungsgemeinschaft (DFG), Project GO 944/4 and EU FET Open RIA Grant no. 766566.

\bibliography{refANECMG}

\end{document}